\documentclass[a4paper,12pt]{article}
\pdfoutput=0
\usepackage{authblk}
\usepackage{amsmath}
\usepackage{diagbox}
\usepackage{graphicx}
\usepackage{subcaption}
\usepackage{booktabs}
\newcommand*\mean[1]{\langle{#1}\rangle}

\title{Mean transverse momentum as a mass composition estimator of cosmic rays}
\author[1]{H. Hedayati Kh.}
\author[2]{H. Hesari\thanks{h.hesari@ipm.ir}}
\affil[1]{Department of Physics, K.N. Toosi University of Technology, P.O. Box 15875-4416, Tehran, Iran}
\affil[2]{School of Particles and Accelerators, Institute for Research in Fundamental Sciences (IPM), P.O. Box 19395-5531, Tehran, Iran}

\date{January 2017}
 
\begin{document}
\maketitle
\begin{abstract}
Determination of mass composition of high energy cosmic rays is one of the greatest challenge in modern astrophysics. All of previous methods for finding the mass composition of primary cosmic rays in a surface array require at least two independent measurements (e.g. muon and electron components) of extensive air showers (EAS). Here a new statistical parameter is introduced which can be used to determine the mass composition of vertical downward cosmic rays in a simple surface array. The main advantage of the introduced parameter is that it does not need two independent measurements and can be used in a simple surface array which does not have muon detectors.
\end{abstract}

\clearpage
\section{Introduction}
Since the beginning of cosmic ray researches, one of the most important questions has been about their nature. After it became clear that most of them are ionized nucleus of common atoms, this question was changed to a question how frequent are individual nucleus in the primary cosmic rays composition. Now, the mass composition of primary cosmic rays is very important for the interpretation of different astrophysical processes and for acceleration mechanisms of ultra high energy cosmic rays.\\ 
Up to energies of about $10^{14}$ eV,  balloon-borne or satellite-borne experiments directly measured the mass composition of primaries with negligible uncertainties \cite{castellina2013astrophysics, Boezio201295}.  At higher energies, direct measurement of mass composition is not possible and mass composition estimation is only possible through extensive air showers which is generated by primary cosmic rays. Propagation of an extensive air shower through atmosphere includes lots of intrinsic fluctuations and so ground based experiments can not access the mass composition of every single primary.\\
Strictly speaking, no EAS experiment measures the mass content of primaries. Instead, they measure one or more of primaries' mass sensitive parameters of EASs generated by primaries. The events can then be interpreted in terms of primary mass by a comparison to air shower simulations using different hadronic interaction models.
Mass composition estimation of primaries (MCEP) can be achieved either in observation of longitudinal development of EAS through atmosphere or simultaneous measurement of the electromagnetic and muonic component of EASs in a surface array. Where, in the former case, people usually determine the EAS  maximum development through atmosphere which is a relatively good mass sensitive parameter.\\
From the EAS maximum development measurements, two observables are derived: $<X_{max}>$, the mean depth of EAS maximum development in $\textrm{g/cm}^2$ and its standard deviation, $\sigma(X_{max})$, which are related to $\ln A$ and $\sigma(\ln A)$, where A is the atomic mass of the primary. EAS simulations shows that proton and iron induced EASs’ maximum development are expected to differ by around 100 $\textrm{g/cm}^2$ . However, unfortunately even extreme mass content like proton and iron have considerable overlap \cite{kampert2012measurements}.
Currently, direct observation of the EAS maximum development is only possible by fluorescence light telescopes. At the moment, there are two active experiments that use fluorescence telescopes for MCEP: The Pierre Auger observatory \cite{abraham2010fluorescence} and the Telescope Array (TA) \cite{matthews2011telescope}. For the Auger Observatory, EAS maximum depth is determined with a resolution of about 25 $\textrm{g/cm}^2$ at low energies decreasing down to about 15 $\textrm{g/cm}^2$ above $10^{18}$ eV. For the latter, the resolution  is better than 40 $\textrm{g/cm}^2$ at energy of $10^{16}$ eV and at higher energies, the accuracy performs better and reaches to 20 $\textrm{g/cm}^2$ \cite{prosin2016results}.\\
Also, non-imaging Cherenkov Telescope can be used for determination of the EAS maximum. It can be deduced either from the measurement of the pulse width at 400 m from the core or from the ratio of the photon densities at two different distances from the shower core \cite{prosin2014tunka}.\\
Recently, the Low Frequency Array, LOFAR \cite{van2013lofar}, a radio telescope consisting of thousands of crossed dipoles, reported a radio measurements of EAS maximum with a high resolution, a mean uncertainty of 16 $\textrm{g/cm}^2$ \cite{buitink2016large}. The situation will be even better with the new proposed installations. For the example of the Square Kilometre Array (SKA) which will be operating in Australia in 2023, it is claimed to reach the resolution of 6 $\textrm{g/cm}^2$ in determination of EAS maximum \cite{huege2016ultimate}. If that is the case, this results in a better mass spectroscopy than ever done before. Although with such a high resolution in determination of $X_{max}$, a far better mass spectroscopy is possible, single event mass determination is nevertheless impossible.\\
Another method for the MCEP is the measurement of particle densities at a surface array. Unlike the observation of EAS maximum development which takes places in a wide range of Earth's atmosphere, surface arrays measure EAS profile only in a single plane and so is more susceptible to the EAS fluctuations. Nevertheless, surface arrays are far more common than telescopes, air Cherenkov detectors and recently developed radio detection techniques. Furthermore, unlike telescopes or air Cherenkov detectors, they have a full duty cycle. So MCEP in surface arrays are still widely used.\\
The most common method for the MCEP in a surface array is the estimation of electron and muon numbers. While the sum of electron and muon numbers at the ground relate to the energy content of an EAS, the ratio of muons number to electrons number is an indication of the primary mass content. The most widely used technique for electron-muon discrimination in a surface array is simultaneous use of unshielded and shielded scintillation detectors (e.g. AGASA \cite{Chiba1992338}, CASA-MIA \cite{Borione1994329}, EAS-TOP \cite{Aglietta1993310}, GRAPES \cite{Gupta2005311}, KASCADE \cite{Antoni2003490}, KASCADE-Grande \cite{Apel2010202}, Maket-ANI \cite{danilova1992ani}, GAMMA \cite{Garyaka2007169}, and Yakutsk \cite{afanasiev1993proceedings}). The Auger observatory uses Cherenkov tanks which also enables a limited muon identification \cite{Abraham200450}.\\
More recently, Canal et al. in their simulations found that the ratio $r_{\mu e}=n_\mu/(E_{\textrm{em}}/0.5\, \textrm{MeV})$ of the muons number to the energy of electromagnetic component of an EAS at the ground level is a good measure for the MCEP. They reported that those vertical EASs with only a value of $r_{\mu e}$ between 0.5 and 3 enable us to reach a $98\%$ efficiency for discrimination of proton from iron primaries \cite{canal2016new}.\\
Another method for the MCEP is the analysis of lateral distribution of charged particles of EASs. Most of lateral distribution functions (LDF) which are in use for energy estimation of EASs have an age parameter which is an estimation of EASs’ maximum development depth. EASs of heavy primaries reach their maximum development depth at higher altitudes than EASs of lighter primaries. So the LDF of heavier primaries are flatter and age parameter has higher values than lighter primaries. However, the precision of this method for the MCEP is lower than other methods (for a review of MCEP methods, see \cite{kampert2012measurements}).\\
In this paper, a new parameter is introduced which enable a good MCEP for vertical EASs. The new parameter can be used in a simple surface array that lacks muon detectors and only measures charged particles in general.
\section{Mean transverse momentom as a mass composition estimator}
\begin{figure}[h!]
    \centering
    \begin{subfigure}[!]{\hsize}
    \centering
        \includegraphics[width=0.6\hsize]{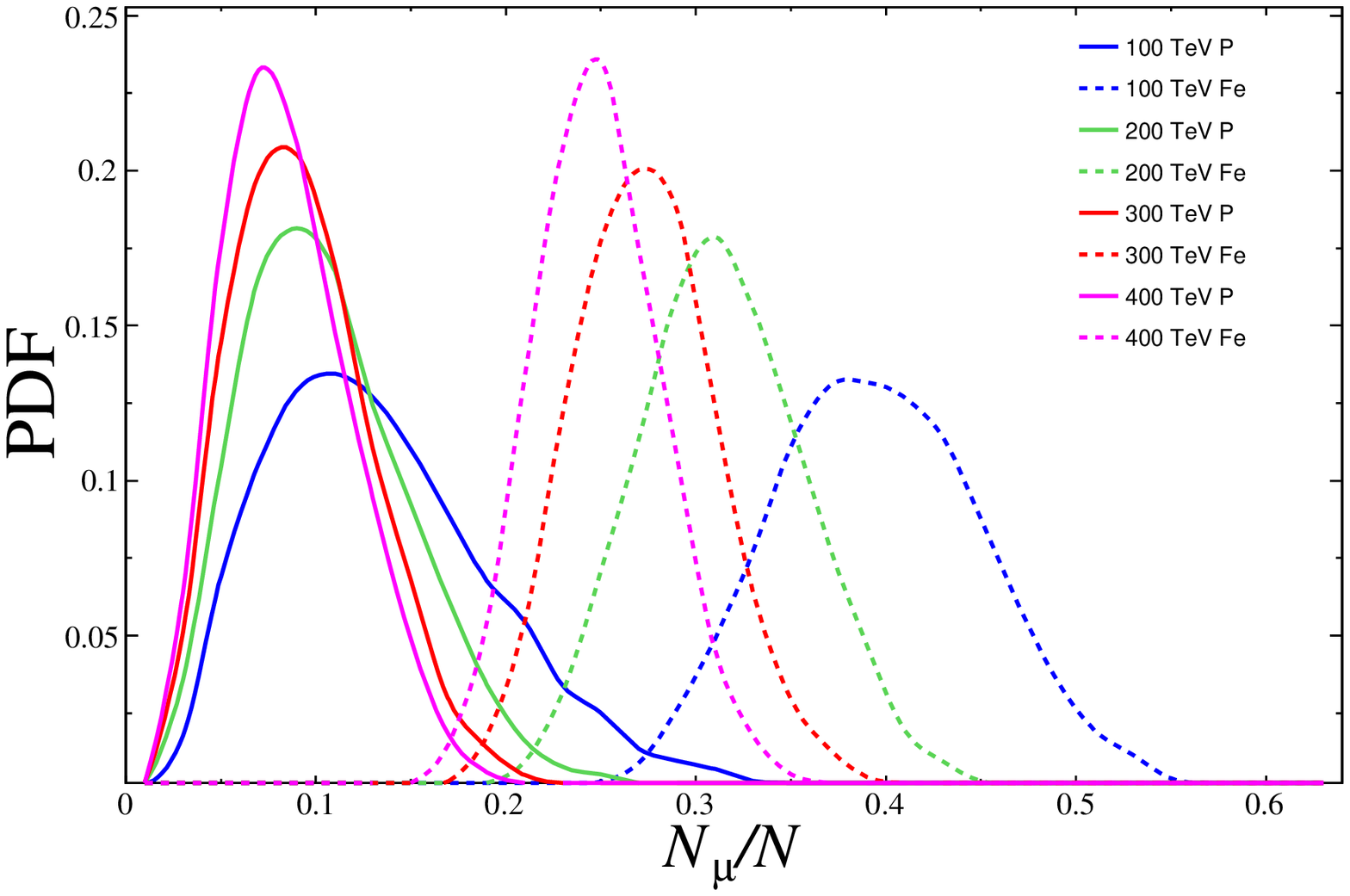}
        \caption{$N_\mu/N$ distributions.}
        \label{NmuN}
    \end{subfigure}
    \begin{subfigure}[!]{\hsize}
    \centering
        \includegraphics[width=0.6\hsize]{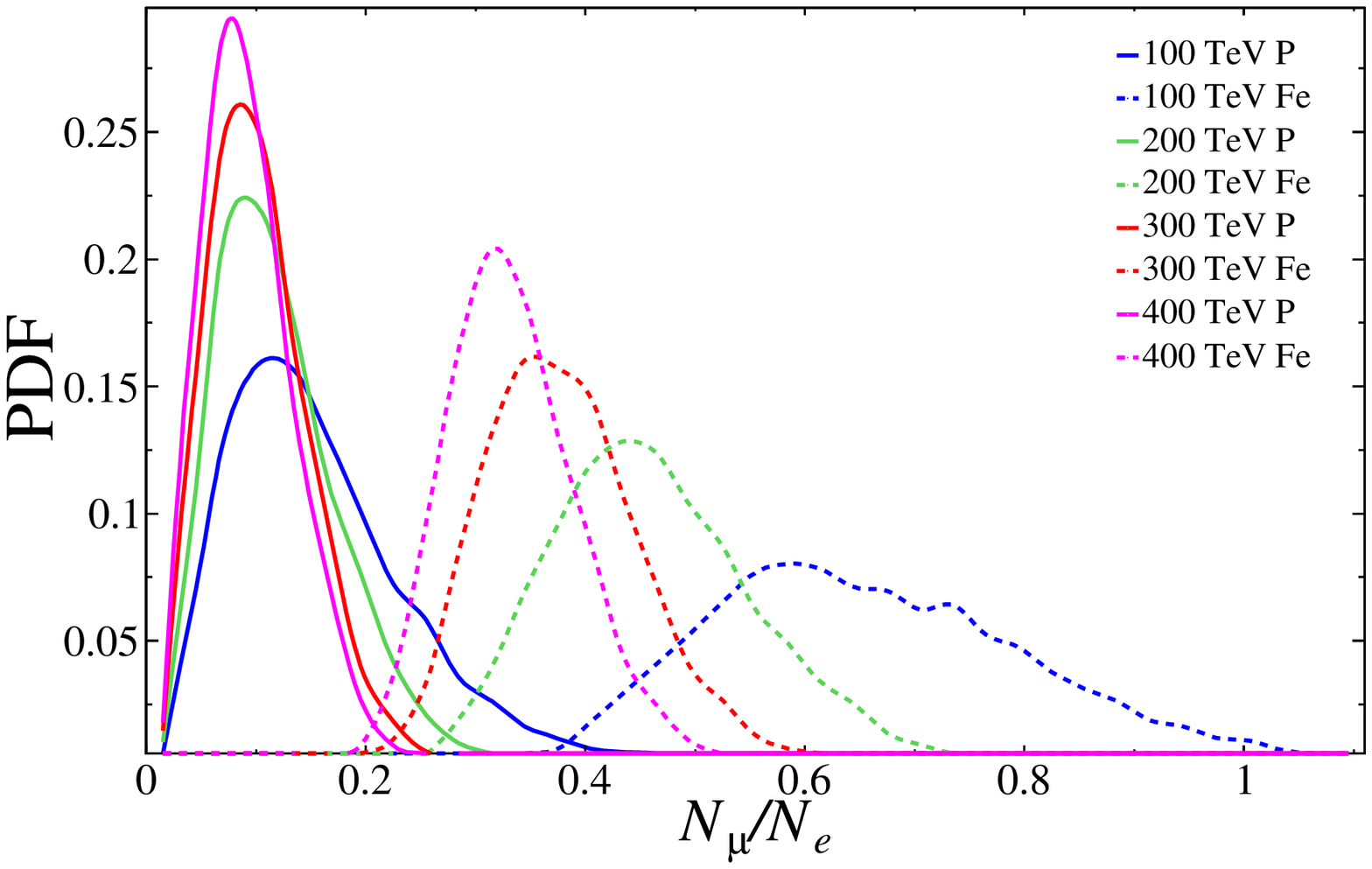}
        \caption{$N_\mu/N_e$ distributions}
        \label{NmuNe}
    \end{subfigure}
    \caption{}\label{NmuNeN}
\end{figure}

From the new results of LHC, it is evident that the more multiplicity of secondary particles of a reaction, the higher transverse momentum per particle of the reaction’s products (e.g. see \cite{alice2013multiplicity}).\\
On the other hand, we conclude from extensive simulations of EASs which conducted by different research groups that the higher the mass composition of primaries, the higher the multiplicity of secondary particles (see chapter 3 of \cite{grieder2010extensive}).\\
The overall result of the above discussion is that when the primary particle is heavier, the generated secondary particles have higher transverse momentum. So if we could determine the transverse momentum per particles ($\mean{P_T}$) of an EAS, we may estimate the primary's rough mass composition.\\
In order to test this hypothesis, some CORSIKA \cite{heck1998corsika} simulated EASs have been generated whose general properties are summarized in table \ref{CORSIKADef}. The primary particles of these EASs are protons or irons. For each type and energy, 10000 EASs have been separately generated.\\
According to table \ref{CORSIKADef}, there are 8 different combinations of types and energies altogether. So the number of all generated EASs are 80000.
\begin{table*}[h!]
\centering
\begin{tabular}{| l | c |} 
 \hline
 Specifications & Values\\ [0.5ex] 
 \hline\hline
 energy of primaries & $E=100,\, 200,\, 300 \textrm{ and }400 \,\textrm{TeV}$\\
 type of primaries & proton and iron
\\
zenith angle of primaries & $\theta=0^\circ$
\\
 geographical longitude & 51 E  \\ 
 geographical latitude & 35 N  \\
 altitude & 1200 m  \\
 earth magnetic field ($B_x$) & $28.1 \,\mu$T  \\
 earth magnetic field ($B_z$) & $38.4 \,\mu$T \\
 low energy hadronic model & Fluka 2011.2b \cite{ferrari2005fluka}  \\
 high energy hadronic model & QGSJETII-04 \cite{ostapchenko2011monte}  \\ [1ex] 
 \hline
\end{tabular}
\caption{EASs' specifications. Other specifications are CORSIKA default values.}
\label{CORSIKADef}
\end{table*}
For the vertical EASs, $\mean{P_T}$ is calculated from the following equation:
\begin{align}
\mean{P_T^e}&=\frac{1}{N_e}\sum_{i=1}^{N_e}{\sqrt{p_{xi}^2+p_{yi}^2}}\\
\mean{P_T^\mu}&=\frac{1}{N_\mu}\sum_{i=1}^{N_\mu}{\sqrt{p_{xi}^2+p_{yi}^2}}\\
\mean{P_T}&=\frac{1}{N}\sum_{i=1}^{N}{\sqrt{p_{xi}^2+p_{yi}^2}}=\frac{N_e}{N}\mean{P_T^e}+\frac{N_\mu}{N}\mean{P_T^\mu}\label{eq.PT}
\end{align}
where $p_{xi}$ and $p_{yi}$ are the components of horizontal momentum of secondary particles.\\
\begin{figure}[h!]
    \centering
    \begin{subfigure}[!]{\hsize}
    \centering
        \includegraphics[width=0.6\hsize]{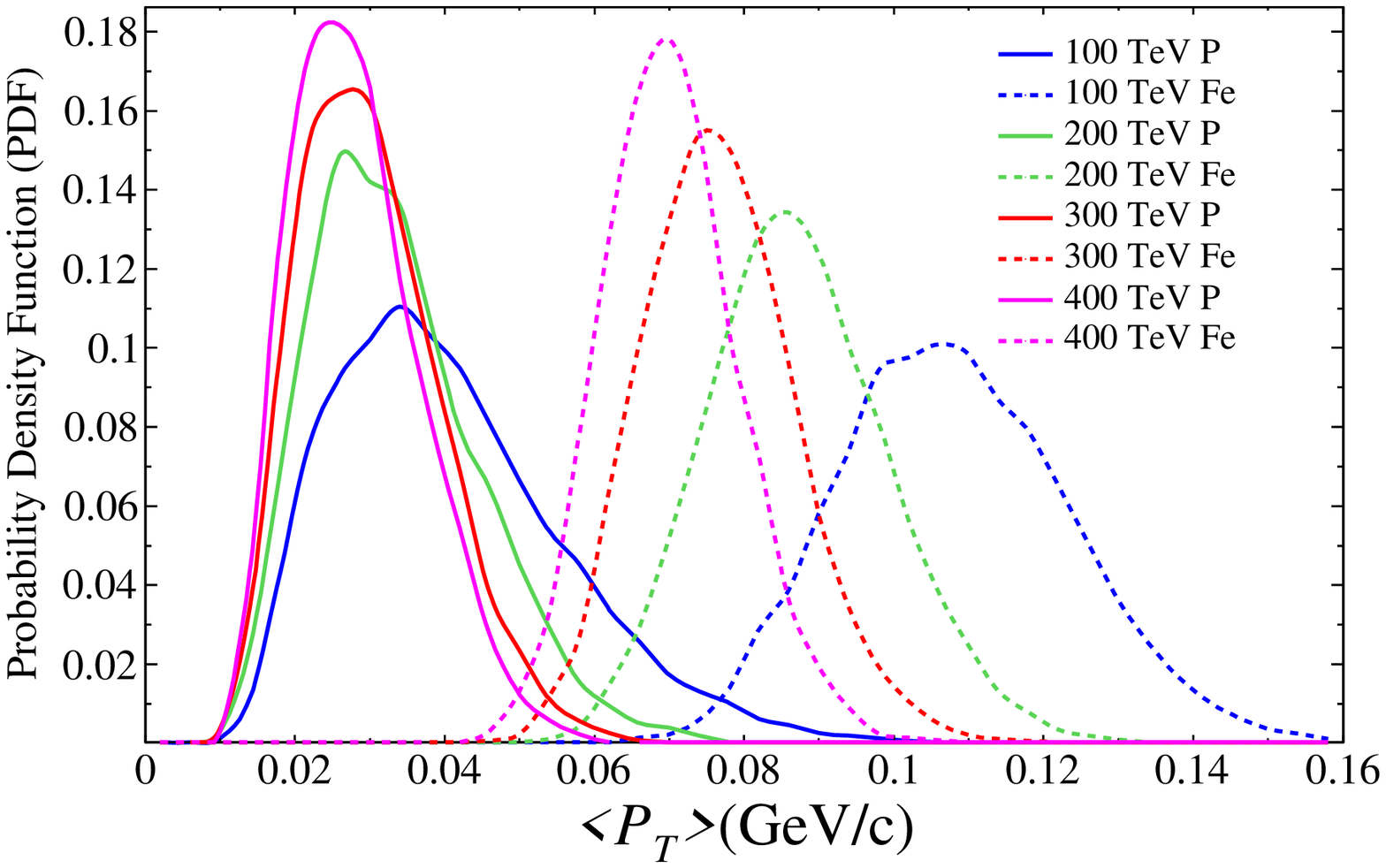}
        \caption{$\mean{P_T}$ distribution for all charged particles.}
        \label{TMP-AC}
    \end{subfigure}
    \begin{subfigure}[!]{\hsize}
    \centering
        \includegraphics[width=0.6\hsize]{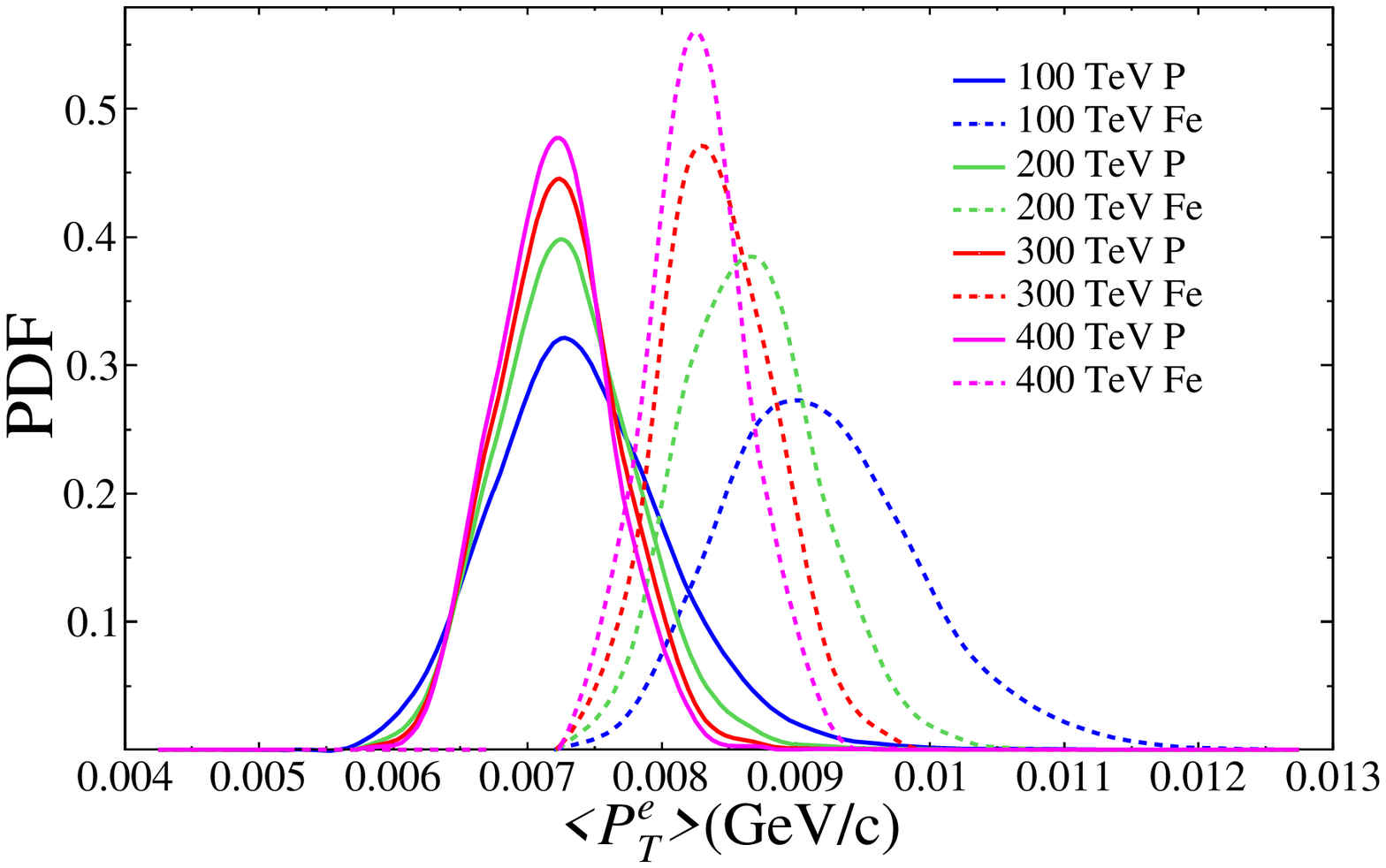}
        \caption{$\mean{P_T^e}$ distribution for electrons and positrons.}
        \label{TMP-E}
    \end{subfigure}
    \begin{subfigure}[!]{\hsize}
    \centering
        \includegraphics[width=0.6\hsize]{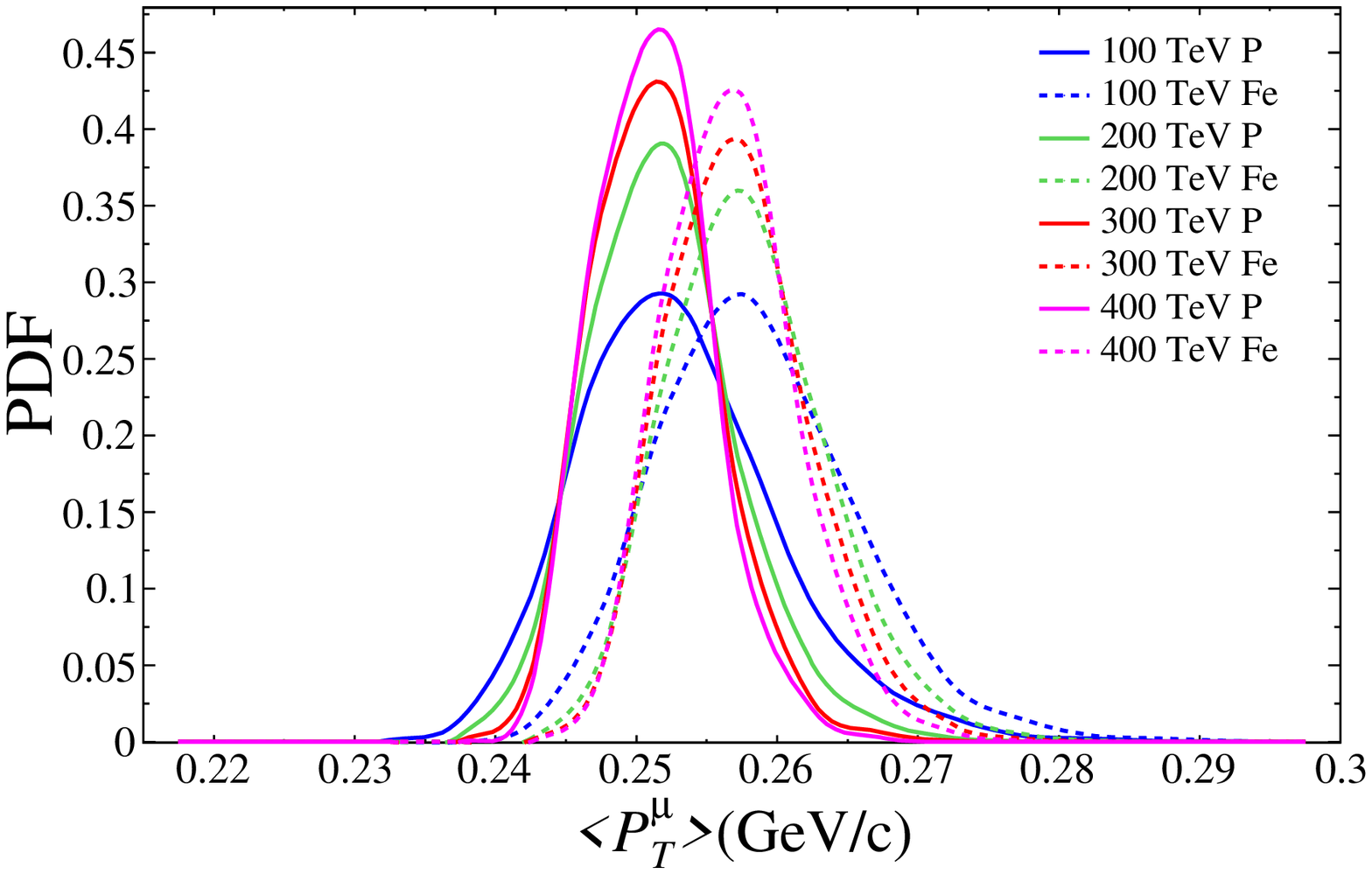}
        \caption{$\mean{P_T^\mu}$ distribution for muons.}
        \label{TMP-M}
    \end{subfigure}
    \caption{The distribution curves represents the $\mean{P_T}$ for secondary particles of vertical simulated EASs. The energies and types of primaries are shown in the margins.}\label{TMP}
\end{figure}
Figure \ref{TMP} shows the distribution of the $\mean{P_T}$ for EASs of table \ref{CORSIKADef} on the ground level. As can be seen in this figure, iron primaries are  distinguished from protons based on their charged secondary particles' $\mean{P_T}$s. Actually $\mean{P_T}$ of charged secondary particles are higher for iron EASs than proton EASs which is consistent with our expectation. Based on part b and c of figure \ref{TMP}, it is evident that the good separation of $\mean{P_T}$ distributions for all charged particles, can not be seen for muonic and electronic components separately. Though at first glance it may seem strange, examining equation \ref{eq.PT} shows that the contribution of coefficeints of $N_e/N$ and $N_\mu/N$ should also be taken into account. It should also be apparent that $\mean{P_T}$ for electronic component is one order of magnitude smaller than $\mean{P_T}$ for muonic component. This difference is due to more interaction of electrons with air nucleus along their path compared to the mouns.\\
\begin{figure}[h!]
    \centering
    \begin{subfigure}[!]{\hsize}
    \centering
        \includegraphics[width=0.6\hsize]{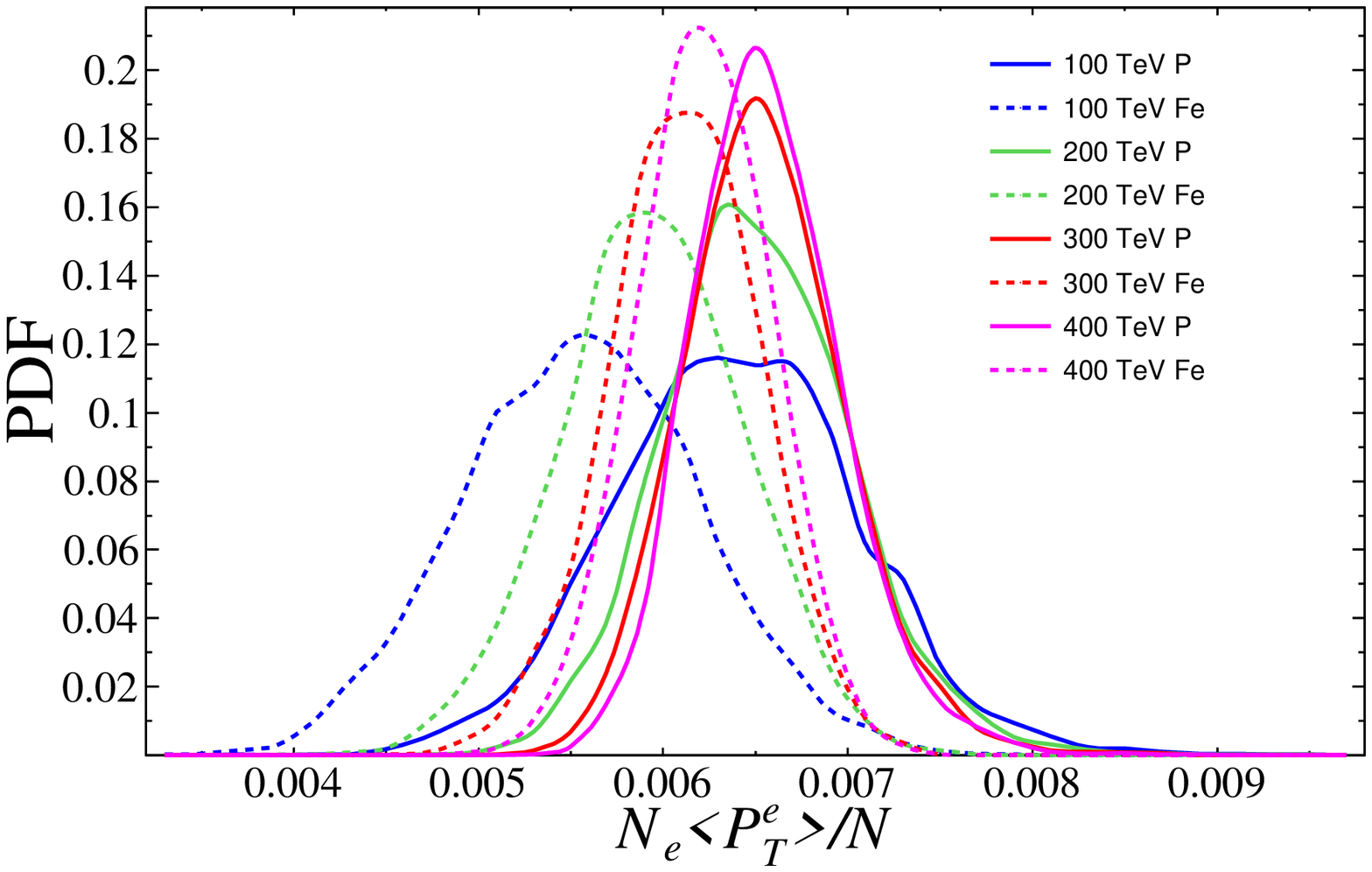}
        \caption{$N_e\mean{P_T^e}/N$ distributions.}
        \label{NePe}
    \end{subfigure}
    \begin{subfigure}[!]{\hsize}
    \centering
        \includegraphics[width=0.6\hsize]{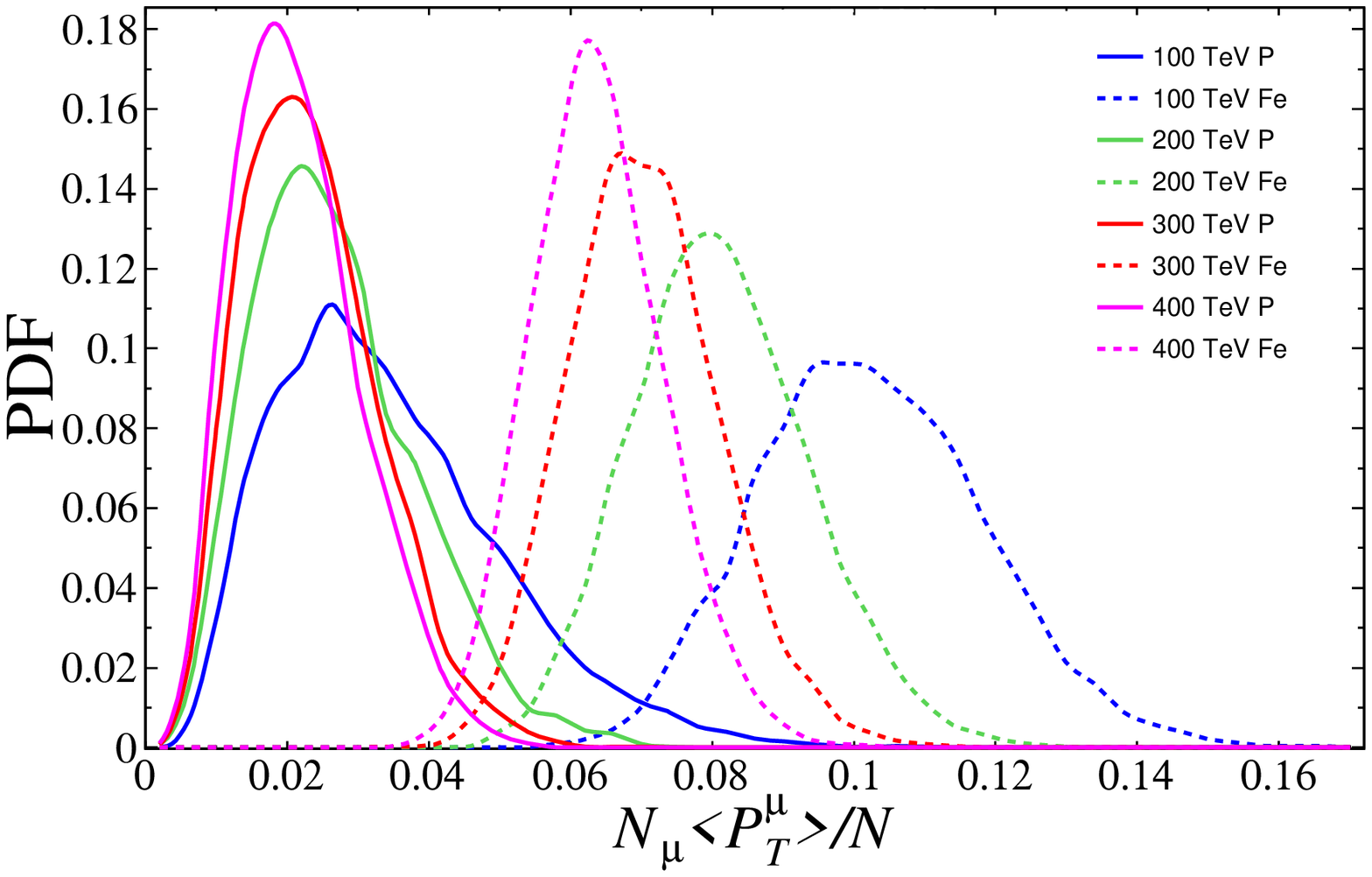}
        \caption{$N_\mu\mean{P_T^\mu}/N$ distributions}
        \label{NmPm}
    \end{subfigure}
    \caption{}\label{NP}
\end{figure}
In order to evaluate the effect of each term of equation \ref{eq.PT} on $\mean{P_T}$ seperately, their distributions have been depicted in figure \ref{NP}. As is evident, $N_\mu\mean{P_T^\mu}/N$ distribution has a far better seperation than $N_e\mean{P_T^e}/N$ distribution. So, the seperation of $\mean{P_T}$ is mainly due to the contribution of muonic term. The electron contribution has negligible effect on the seperation.\\
Next, we sould evaluate the effect of $N_\mu/N$ on the seperation. Figure \ref{NmuN} shows the results of the $N_\mu/N$ distribution. A comparison of this figure and figure \ref{TMP-M}, convince us that the $N_\mu/N$ coefficient has the most important role in the MCEP of the $\mean{P_T}$. Figure \ref{NmuNe} also shows the results of $N_\mu/N_e$ for reference.\\
Up to now, we only have done  a qualitative investigation of the MCEP parameters. However, a quantitative comarison will be more informative. Also, the precisions have not been investigated with respect to the energy of EASs.\\
In order to study the resolution of $\mean{P_T}$ and $N_\mu/N$ for the MCEP, we introduce the resolution coefficient with the following equation:
\begin{equation}
R=\frac{\lvert m_{\rm{P}}-m_{\rm{Fe}}\rvert}{\sigma_{\rm{P}}+\sigma_{\rm{Fe}}}
\end{equation}
where $m$ is the mean value of a distribution and $\sigma$ is the standard deviation. Actually, resolution coefficietnt $R$ is a measure of distance between two distributions' averages as well as the narrowness of the distributions. The merit of such a dimensionless parameter is that it checks not only the sepration of the two distributions' averages but also the mixing amount of them.\\
As can be seen in table \ref{RCoePNmu}, both of $\mean{P_T}$ and $N_\mu/N$ have almost the same precision. Altough, $\mean{P_T}$ has slightly less precision, it has an obvious advantage: It only uses the information of all charged particles and do not need muon specilized detectors. Also according to table \ref{RCoePNmu}, $\mean{P_T}$ and $N_\mu/N$ have better results than the $N_\mu/Ne$.
\begin{table}[!h]
\centering
\begin{tabular}{ccccc}
\toprule
$R$ & 100 TeV & 200 TeV & 300 TeV & 400 TeV\\
\midrule
$\mean{P_T}$ & 2.17 & 2.29 & 2.35 & 2.37\\
\midrule
$N_\mu/N$ & 2.21 & 2.32 & 2.37 & 2.38\\
\midrule
$N_\mu/N_e$ & 2.06 & 2.18 & 2.26 & 2.30\\
\midrule
$\mean{d_T}$ & 2.02 & 2.05 & 2.079 & 2.066\\
\bottomrule
\end{tabular}
\caption{Resolution coefficients $R$ of MCEP parameters.}
\label{RCoePNmu}
\end{table}
\section{Mean transverse distance from axis}
Although, $\mean{P_T}$ is a useful parameter for the MCEP, it can not be directly obtained by the current surface arrays in use (A surface array of hodoscopes is probably appropriate for this purpose). Therefore, we should find a new parameter which has some correlation with the above-mensioned parameters and can be obtained from the information provided by a common surface array.\\
Although the propagation of an EAS in the atmosphere has lots of stochastic fluctuations, obviously when the $\mean{P_T}$ of an EAS is large, its secondary particles have a higher chance to be found at larger distances from axis. So, a new variable of an EAS which is correlated to $\mean{P_T}$ is the mean transverse distance of particles from the axis ($\mean{d_T}$):
\begin{equation}
\mean{d_T}=\frac{1}{N}\sum_{i=1}^N{d_i},
\end{equation}
where $d_i$ is the distance of $i$th secondary particle from the axis.\\
\begin{center}
\begin{tabular}{|l|ll|}
\hline
\backslashbox{Energy}{Type} & proton & iron\\
\hline
\hline
 100 TeV & 0.98 & 0.97\\
200 TeV & 0.97 & 0.98 \\ 
 300 TeV & 0.97 &  0.98\\
400 TeV & 0.97& 0.98 \\
\hline
\end{tabular}
\captionof{table}{Correlation coefficient between $\mean{P_T}$ and $\mean{d_T}$}\label{CorCoe}
\end{center}
Table \ref{CorCoe} shows the correlation coefficient between $\mean{P_T}$ and $\mean{d_T}$ for the EASs of table \ref{CORSIKADef}. This correlation coefficient is given by the following equation:
\begin{align}
\rho_{Pd} &=\frac{cov(\mean{P_T},\mean{d_T})}{\sigma_P \sigma_d},\\
cov(\mean{P_T},\mean{d_T})&=E[(\mean{P_T}-\mu_P)(\mean{d_T}-\mu_d)] \nonumber \\
&=\frac{1}{N_{EASs}}\sum_{i=1}^{N_{EASs}} (\mean{P_T}_i-\mu_P)(\mean{d_T}_i-\mu_d),\nonumber \\
\mu_P &=E[\mean{P_T}]=\frac{1}{N_{EASs}}\sum_{i=1}^{N_{EASs}} \mean{P_T}_i, \nonumber \\
\mu_d &=E[\mean{d_T}]=\frac{1}{N_{EASs}}\sum_{i=1}^{N_{EASs}} \mean{d_T}_i, \nonumber \\
\sigma_P^2&=\frac{1}{N_{EASs}}\sum_{i=1}^{N_{EASs}}(\mean{P_T}_i-\mu_P)^2, \nonumber \\
\sigma_d^2&=\frac{1}{N_{EASs}}\sum_{i=1}^{N_{EASs}}(\mean{d_T}_i-\mu_d)^2  \nonumber
\end{align}
where $\mean{P_T}_i$ and $\mean{d_T}_i$ are the mean transverse momentum and mean transverse distance of $i$th EAS, the $E$ means the ensemble average over all the EASs, $\mu_P$ and $\mu_d$ are the ensemble average of $\mean{P_T}$ and $\mean{d_T}$ of all EASs, respectively and $N_{EASs}$ parameter is the total number of EASs that is here 10000.\\
\begin{figure}[ht!]
    \centering
    \includegraphics[width=0.6\hsize]{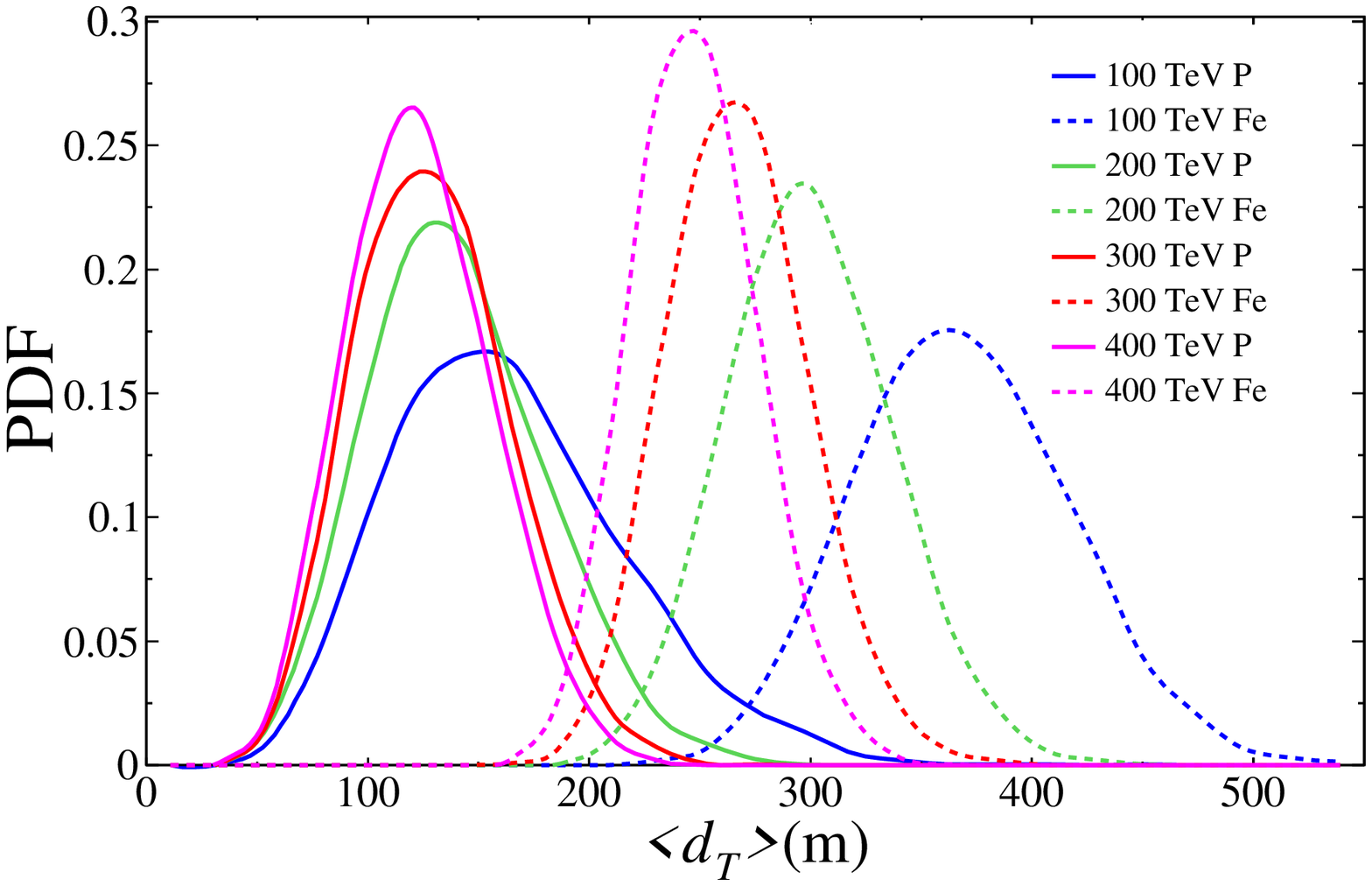}
	\caption{$\mean{d_T}$ distribution for all charged particles.}
    \label{dT}
\end{figure}
As is evident from table \ref{CorCoe}, the correlation between $\mean{P_T}$ and $\mean{d_T}$ is nearly perfect. So $\mean{d_T}$ can be safely used instead of $\mean{P_T}$ for MCEP. Figure \ref{dT} shows the distribution of $\mean{d_T}$. It is apparent from this figure that $\mean{d_T}$ is a suitable parameter for the MCEP, as it was anticipated.\\
As we can see from figure \ref{dT}, $\mean{d_T}$ can also discriminate between P and Fe with a relatively high resolution. Another interesting feature in figure \ref{dT} is that the distribution of $\mean{d_T}$ becomes sharper with increasing energy of the primaries (especially for the irons). Also the $\mean{d_T}$ approaches to lower (but distinct) values for irons than protons.\\
Table \ref{RCoePNmu} also shows the $R$ coefficient for the distributions of figure \ref{dT}. As can be seen from this table, this parameter have better results in the higher energies. 
\section{Conclusions}
In this paper a new parameter has been introduced which can be applied for the mass composition estimation of vertical primary cosmic rays. This parameter is the transverse momentum per secondary particles of an EAS, $\mean{P_T}$. In an obvious manner, we concluded that $\mean{P_T}$ can be reduced to $N_\mu/N$ parameter for the MCEP with a negligible loss of information. It is also shown that a related parameter to $\mean{P_T}$ which can be directly measured in a simple surface array without muonic specialized detectors, is the mean transverse distance from axis, $\mean{d_T}$. It is also evident that the distributions of these parameters are sharper in higher energies.\\


\begin{thebibliography}{10}

\bibitem{abraham2010fluorescence}
J~Abraham, P~Abreu, M~Aglietta, C~Aguirre, EJ~Ahn, D~Allard, I~Allekotte,
  J~Allen, P~Allison, J~Alvarez-Muniz, et~al.
\newblock The fluorescence detector of the pierre auger observatory.
\newblock {\em Nuclear Instruments and Methods in Physics Research Section A:
  Accelerators, Spectrometers, Detectors and Associated Equipment},
  620(2):227--251, 2010.

\bibitem{Abraham200450}
J.~Abraham, M.~Aglietta, I.C. Aguirre, M.~Albrow, D.~Allard, I.~Allekotte,
  P.~Allison, J.~Alvarez Muñiz, M.G. do~Amaral, M.~Ambrosio, L.~Anchordoqui,
  R.~Andrews, M.~Anguiano, J.C. dos Anjos, C.~Aramo, S.~Argiro, K.~Arisaka,
  J.C. Arteaga, S.~Atulugama, M.~Ave, G.~Avila, R.~Baggio, X.~Bai, A.F.
  Barbosa, H.M.J. Barbosa, D.~Barnhill, S.L.C. Barroso, P.~Bauleo, J.~Beatty,
  T.~Beau, K.-H. Becker, J.A. Bellido, P.~Bello, T.~Bergmann, E.~Berman,
  X.~Bertou, P.~Biermann, P.~Billoir, R.~Biral, H.~Bluemer, M.~Bohacova,
  E.~Bollmann, C.~Bonifazi, M.~Boratav, A.~Boselli, J.~Brack, J.-M. Brunet,
  H.~Bui-Duc, V.~Cabrera, D.V. Camin, J.-N. Capdevielle, A.~Carreño,
  N.~Cartiglia, R.~Caruso, L.A. de~Carvalho, S.~Casanova, E.~Casimiro,
  A.~Castellina, J.~Castro, P.W. Cattaneo, L.~Cazon, R.~Cester, N.~Chávez,
  D.~Cheam, A.~Chiavassa, J.A. Chinellato, M.~Chiosso, A.~Chou, J.~Chye,
  A.~Cillis, B.~Civit, D.~Claes, P.D.J. Clark, R.W. Clay, F.~Cohen, A.~Cordero,
  A.~Cordier, E.~Cormier, J.~Cotzomi, U.~Cotti, S.~Coutu, C.E. Covault,
  A.~Creusot, J.W. Cronin, M.~Cuautle, S.~Dagoret-Campagne, T.~Dang-Quang,
  P.~Da Silva, J.~Darling, P.~Darriulat, K.~Daumiller, B.R. Dawson,
  L.~de~Bruijn, A.~De Capoa, M.A.L. de~Oliveira, V.~de~Souza, A.~Della Selva,
  O.~Deligny, J.C. Diaz, C.~Dobrigkeit, J.C. D'Olivo, A.~Dorofeev, M.T. Dova,
  A.~Dye, M.A. DuVernois, R.~Engel, L.N. Epele, P.~Eschstruth, C.O. Escobar,
  A.~Etchegoyen, P.~Facal~San Luis, A.C. Fauth, N.~Fazzini, A.~Fernández,
  A.M.J. Ferrero, B.~Fick, A.~Filevich, A.~Filipčič, R.~Fonte, W.~Fulgione,
  E.~Gámez, B.~Garcia, C.A. Garcia, H.~Geenen, H.~Gemmeke, C.~Germain-Renaud,
  P.L. Ghia, K.~Gibbs, M.~Giller, J.~Gitto, H.~Glass, M.~Gómez Berisso,
  P.F.~Gomez Vitale, J.~González, J.~González, D.~Gora, A.~Goodwin,
  P.~Gouffon, V.~Grassi, A.F. Grillo, C.~Grunfeld, J.~Grygar, F.~Guarino,
  G.~Guedes, C.~Guerard, R.~Gumbsheimer, J.L. Harton, F.~Hasenbalg, D.~Heck,
  J.M. Hernández, D.~Hoffer, C.~Hojvat, P.~Homola, M.~Horvat, M.~Hrabovsky,
  A.~Insolia, S.~Jaminion, Y.~Jerónimo, L.~Jiang, M.~Kaducak, K.-H. Kampert,
  B.~Keilhauer, E.~Kemp, H.~Klages, M.~Kleifges, J.~Kleinfeller, J.~Knapp,
  A.~Kopmann, N.~Kunka, M.~Kutschera, C.~Lachaud, M.~Lapolla,
  A.~Letessier-Selvon, I.~Lhenry-Yvon, J.~Lloyd-Evans, R.~López, A.~Lopez
  Aguera, M.~Lucano, R.~Luna, Y.~Ma, M.E. Manceñido, P.F. Manfredi,
  L.~Manhaes, P.~Mantsch, A.G. Mariazzi, M.J. Markus, G.~Martin, O.~Martineau,
  J.~Martinez, N.~Martinez, O.~Martı́nez, H.-J. Mathes, J.A.J. Matthews,
  J.~Matthews, G.~Matthiae, E.~Marques, P.~Matussek, G.~Maurin, D.~Maurizio,
  P.~Mazur, T.~McCauley, M.~McEwen, R.R. McNeil, C.~Medina, M.C. Medina,
  G.~Medina-Tanco, D.~Melo, M.~Melocchi, E.~Menichetti, A.~Menshikov, F.~Meyer,
  R.~Meyhandan, J.C. Meza, G.~Miele, W.~Miller, M.~Mohammed,
  D.~Monnier-Ragaigne, C.~Morello, E.~Moreno, M.~Mostafa, R.~Mussa, H.~Nassini,
  G.~Navarra, L.~Nellen, F.~Nerling, C.~Newman-Holmes, D.~Nicotra, S.~Nigro,
  D.~Nitz, H.~Nogima, D.~Nosek, M.~Nuñez, T.~Ohnuki, A.~Olinto,
  S.~Ostaptchenko, M.~Palatka, G.~Parente, E.~Parizot, E.H. Pasaye,
  N.~Pastrone, M.~Patel, T.~Paul, I.~Pedraza, J.~Pekala, R.~Pelayo, I.M. Pepe,
  A.~Pérez-Lorenzana, L.~Perrone, N.~Peshman, S.~Petrera, P.~Petrinca,
  D.~Pham-Ngoc, P.~Pham-Trung, T.~Pierog, O.~Pisanti, N.~Playez, E.~Ponce, T.A.
  Porter, L.~Prado Junior, P.~Privitera, M.~Prouza, C.L. Pryke, J.B. Rafert,
  G.~Raia, S.~Ranchon, L.~Ratti, D.~Ravignani, V.~Re, H.C. Reis, S.~Reucroft,
  B.~Revenue, M.~Richter, J.~Ridky, A.~Risi, M.~Risse, V.~Rizi, M.D. Roberts,
  C.~Robledo, G.~Rodriguez, J.~Rodriquez, J.~Rodriquez Martino, S.~Román,
  L.~Rosa, M.~Roth, A.C. Rovero, H.~Salazar, G.~Salina, F.~Sanchez,
  M.~Santander, L.G. dos Santos, R.~Sato, P.~Schovanek, V.~Scherini, S.J.
  Sciutto, G.~Sequeiros-Haddad, R.C. Shellard, E.~Shibuya, F.V. Siguas,
  W.~Slater, N.~Smetniansky-De Grande, K.~Smith, G.R. Snow, P.~Sommers,
  C.~Song, H.~Spinka, F.~Suarez, T.~Suomijärvi, D.~Supanitsky, J.~Swain,
  Z.~Szadkowski, A.~Tamashiro, G.J. Thornton, T.~Thouw, R.~Ticona, W.~Tkaczyk,
  C.J.~Todero Peixoto, A.~Tripathi, G.~Tristram, M.~Trombley,
  D.~Tscherniakhovski, P.~Tuckey, V.~Tunnicliffe, M.~Urban, C.~Uribe Estrada,
  J.F. Valdés, A.~Vargas, C.~Vargas, R.~Vazquez, D.~Veberič, A.~Veiga,
  A.~Velarde, F.~Vernotte, V.~Verzi, M.~Videla, C.~Vigorito, L.M. Villaseñor,
  M.~Vlcek, L.~Voyvodic, T.~Vo-Van, T.~Waldenmaier, P.~Walker, D.~Warner, A.A.
  Watson, Ch. Wiebusch, G.~Wieczorek, B.~Wilczynska, H.~Wilczynski, N.R. Wild,
  T.~Yamamoto, E.~Zas, D.~Zavrtanik, M.~Zavrtanik, A.~Zepeda, C.~Zhang, and
  Q.~Zhu.
\newblock Properties and performance of the prototype instrument for the pierre
  auger observatory.
\newblock {\em Nuclear Instruments and Methods in Physics Research Section A:
  Accelerators, Spectrometers, Detectors and Associated Equipment},
  523(1–2):50 -- 95, 2004.

\bibitem{afanasiev1993proceedings}
B.~N. Afanasiev and {\it et al.}~[Yakutsk~Collaboration].
\newblock Proceedings of the tokyo workshop on techniques for the study of the
  extremely high energy cosmic rays.
\newblock page~35, 1993.

\bibitem{Aglietta1993310}
M.~Aglietta, B.~Alessandro, P.~Antonioli, F.~Arneodo, L.~Bergamasco,
  A.~Campos~Fauth, C.~Castagnoli, A.~Castellina, C.~Cattadori, A.~Chiavassa,
  G.~Cini, B.~D'Ettorre~Piazzoli, G.~Di~Sciascio, W.~Fulgione, P.~Galeotti,
  P.L. Ghia, M.~Iacovacci, G.~Mannocchi, C.~Morello, G.~Navarra, L.~Riccati,
  O.~Saavedra, G.C. Trinchero, P.~Vallania, and S.~Vernetto.
\newblock Uhe cosmic ray event reconstruction by the electromagnetic detector
  of eas-top.
\newblock {\em Nuclear Inst. and Methods in Physics Research, A},
  336(1-2):310--321, 1993.
\newblock cited By 70.

\bibitem{Antoni2003490}
T.~Antoni, W.D. Apel, F.~Badea, K.~Bekk, A.~Bercuci, H.~Blümer, H.~Bozdog,
  I.M. Brancus, C.~Büttner, A.~Chilingarian, K.~Daumiller, P.~Doll, J.~Engler,
  F.~Feßler, H.J. Gils, R.~Glasstetter, R.~Haeusler, A.~Haungs, D.~Heck, J.R.
  Hörandel, A.~Iwan, K.-H. Kampert, H.O. Klages, G.~Maier, H.J. Mathes, H.J.
  Mayer, J.~Milke, M.~Müller, R.~Obenland, J.~Oehlschläger, S.~Ostapchenko,
  M.~Petcu, H.~Rebel, M.~Risse, M.~Roth, G.~Schatz, H.~Schieler, J.~Scholz,
  T.~Thouw, H.~Ulrich, A.~Vardanyan, J.~Weber, A.~Weindl, J.~Wentz, J.~Wochele,
  J.~Zabierowski, and S.~Zagromski.
\newblock The cosmic-ray experiment kascade.
\newblock {\em Nuclear Instruments and Methods in Physics Research, Section A:
  Accelerators, Spectrometers, Detectors and Associated Equipment},
  513(3):490--510, 2003.
\newblock cited By 200.

\bibitem{Apel2010202}
W.D. Apel, J.C. Arteaga, A.F. Badea, K.~Bekk, M.~Bertaina, J.~Blmer, H.~Bozdog,
  I.M. Brancus, P.~Buchholz, E.~Cantoni, A.~Chiavassa, F.~Cossavella,
  K.~Daumiller, V.~De~Souza, F.~Di~Pierro, P.~Doll, R.~Engel, J.~Engler,
  M.~Finger, D.~Fuhrmann, P.L. Ghia, H.J. Gils, R.~Glasstetter, C.~Grupen,
  A.~Haungs, D.~Heck, J.R. Hrandel, T.~Huege, P.G. Isar, K.-H. Kampert,
  D.~Kang, D.~Kickelbick, H.O. Klages, K.~Link, P.~Uczak, M.~Ludwig, H.J.
  Mathes, H.J. Mayer, M.~Melissas, J.~Milke, B.~Mitrica, C.~Morello,
  G.~Navarra, S.~Nehls, J.~Oehlschlger, S.~Ostapchenko, S.~Over, N.~Palmieri,
  M.~Petcu, T.~Pierog, H.~Rebel, M.~Roth, H.~Schieler, F.~Schrder, O.~Sima,
  M.~Stmpert, G.~Toma, G.C. Trinchero, H.~Ulrich, A.~Weindl, J.~Wochele,
  M.~Wommer, and J.~Zabierowski.
\newblock The kascade-grande experiment.
\newblock {\em Nuclear Instruments and Methods in Physics Research, Section A:
  Accelerators, Spectrometers, Detectors and Associated Equipment},
  620(2-3):202--216, 2010.
\newblock cited By 93.

\bibitem{Boezio201295}
Mirko Boezio and Emiliano Mocchiutti.
\newblock Chemical composition of galactic cosmic rays with space experiments.
\newblock {\em Astroparticle Physics}, 39–40:95 -- 108, 2012.
\newblock Cosmic Rays Topical Issue.

\bibitem{Borione1994329}
A.~Borione, C.E. Covault, J.W. Cronin, B.E. Fick, K.G. Gibbs, H.A. Krimm, N.C.
  Mascarenhas, T.A. McKay, D.~Müller, B.J. Newport, R.A. Ong, L.J. Rosenberg,
  H.~Sanders, M.~Catanese, D.~Ciampa, K.D. Green, J.~Kolodziejczak,
  J.~Matthews, D.~Nitz, D.~Sinclair, and J.C. van~der Velde.
\newblock A large air shower array to search for astrophysical sources emitting
  γ-rays with energies ≥1014 ev.
\newblock {\em Nuclear Inst. and Methods in Physics Research, A},
  346(1-2):329--352, 1994.
\newblock cited By 69.

\bibitem{buitink2016large}
S~Buitink, A~Corstanje, H~Falcke, JR~H{\"o}randel, T~Huege, A~Nelles,
  JP~Rachen, L~Rossetto, P~Schellart, O~Scholten, et~al.
\newblock A large light-mass component of cosmic rays at 1017--1017.5
  electronvolts from radio observations.
\newblock {\em Nature}, 531(7592):70--73, 2016.

\bibitem{canal2016new}
CA~Garc{\'\i}a Canal, JI~Illana, M~Masip, and SJ~Sciutto.
\newblock A new observable in extensive air showers.
\newblock {\em Astroparticle Physics}, 2016.

\bibitem{castellina2013astrophysics}
Antonella Castellina and Fiorenza Donato.
\newblock Astrophysics of galactic charged cosmic rays.
\newblock {\em Planets, Stars and Stellar Systems: Volume 5: Galactic Structure
  and Stellar Populations}, pages 725--788, 2013.

\bibitem{Chiba1992338}
N.~Chiba, K.~Hashimoto, N.~Hayashida, K.~Honda, M.~Honda, N.~Inoue,
  F.~Kakimoto, K.~Kamata, S.~Kawaguchi, N.~Kawasumi, Y.~Matsubara, K.~Murakami,
  M.~Nagano, S.~Ogio, H.~Ohoka, To. Saito, Y.~Sakuma, I.~Tsushima, M.~Teshima,
  T.~Umezawa, S.~Yoshida, and H.~Yoshii.
\newblock Akeno giant air shower array (agasa) covering 100 km2 area.
\newblock {\em Nuclear Inst. and Methods in Physics Research, A},
  311(1-2):338--349, 1992.
\newblock cited By 95.

\bibitem{alice2013multiplicity}
ALICE collaboration et~al.
\newblock Multiplicity dependence of the average transverse momentum in pp,
  p--pb, and pb--pb collisions at the lhc.
\newblock {\em Physics Letters B}, 727(4):371--380, 2013.

\bibitem{danilova1992ani}
TV~Danilova, EA~Danilova, AB~Ezlikin, NM~Nikolskaia, SJ~Nikolsky, VA~Romachin,
  SA~Slavatinsky, BV~Subbotin, EJ~Tukich, KM~Avakian, et~al.
\newblock The ani experiment: on the investigation of interactions from hadrons
  and nuclei in the energy range 103--105 tev.
\newblock {\em Nuclear Instruments and Methods in Physics Research Section A:
  Accelerators, Spectrometers, Detectors and Associated Equipment},
  323(1-2):104--107, 1992.

\bibitem{ferrari2005fluka}
Alfredo Ferrari, Paola~R Sala, Alberto Fasso, and Johannes Ranft.
\newblock Fluka: A multi-particle transport code (program version 2005).
\newblock Technical report, 2005.

\bibitem{Garyaka2007169}
A.P. Garyaka, R.M. Martirosov, S.V. Ter-Antonyan, N.~Nikolskaya, Y.A. Gallant,
  L.~Jones, and J.~Procureur.
\newblock Rigidity-dependent cosmic ray energy spectra in the knee region
  obtained with the gamma experiment.
\newblock {\em Astroparticle Physics}, 28(2):169--181, 2007.
\newblock cited By 22.

\bibitem{grieder2010extensive}
Peter~Karl Grieder.
\newblock {\em Extensive Air Showers}.
\newblock Springer, 2010.

\bibitem{Gupta2005311}
S.K. Gupta, Y.~Aikawa, N.V. Gopalakrishnan, Y.~Hayashi, N.~Ikeda, N.~Ito,
  A.~Jain, A.V. John, S.~Karthikeyan, S.~Kawakami, T.~Matsuyama, D.K. Mohanty,
  P.K. Mohanty, S.D. Morris, T.~Nonaka, A.~Oshima, B.S. Rao, K.C. Ravindran,
  M.~Sasano, K.~Sivaprasad, B.V. Sreekantan, H.~Tanaka, S.C. Tonwar,
  K.~Viswanathan, and T.~Yoshikoshi.
\newblock Grapes-3 - a high-density air shower array for studies on the
  structure in the cosmic-ray energy spectrum near the knee.
\newblock {\em Nuclear Instruments and Methods in Physics Research, Section A:
  Accelerators, Spectrometers, Detectors and Associated Equipment},
  540(2-3):311--323, 2005.
\newblock cited By 39.

\bibitem{heck1998corsika}
Dieter Heck, G~Schatz, J~Knapp, T~Thouw, and JN~Capdevielle.
\newblock Corsika: A monte carlo code to simulate extensive air showers.
\newblock Technical report, 1998.

\bibitem{huege2016ultimate}
Tim Huege, Justin~D Bray, Stijn Buitink, David Butler, Richard Dallier, Ron~D
  Ekers, Torsten En{\ss}lin, Heino Falcke, Andreas Haungs, Clancy~W James,
  et~al.
\newblock Ultimate precision in cosmic-ray radio detection---the ska.
\newblock {\em arXiv preprint arXiv:1608.08869}, 2016.

\bibitem{kampert2012measurements}
Karl-Heinz Kampert and Michael Unger.
\newblock Measurements of the cosmic ray composition with air shower
  experiments.
\newblock {\em Astroparticle Physics}, 35(10):660--678, 2012.

\bibitem{matthews2011telescope}
JN~Matthews, C~Jui, Pierre Sokolsky, M~Fukushima, G~Thomson, H~Sagawa, and
  S~Ogio.
\newblock The telescope array experiment.
\newblock In {\em Institute of High Energy Physics}, 2011.

\bibitem{ostapchenko2011monte}
Sergey Ostapchenko.
\newblock Monte carlo treatment of hadronic interactions in enhanced pomeron
  scheme: Qgsjet-ii model.
\newblock {\em Physical Review D}, 83(1):014018, 2011.

\bibitem{prosin2014tunka}
VV~Prosin, SF~Berezhnev, NM~Budnev, A~Chiavassa, OA~Chvalaev, OA~Gress,
  AN~Dyachok, SN~Epimakhov, NI~Karpov, NN~Kalmykov, et~al.
\newblock Tunka-133: Results of 3 year operation.
\newblock {\em Nuclear Instruments and Methods in Physics Research Section A:
  Accelerators, Spectrometers, Detectors and Associated Equipment},
  756:94--101, 2014.

\bibitem{prosin2016results}
VV~Prosin, NM~Budnev, A~Chiavassa, AN~Dyachok, SN~Epimakhov, F~Fenu, Yu~A
  Fomin, OA~Gress, TI~Gress, NN~Kalmykov, et~al.
\newblock Results and perspectives of cosmic ray mass composition studies with
  eas arrays in the tunka valley.
\newblock In {\em Journal of Physics: Conference Series}, volume 718, page
  052031. IOP Publishing, 2016.

\bibitem{van2013lofar}
MP~Van~Haarlem, MW~Wise, AW~Gunst, George Heald, JP~McKean, JWT Hessels,
  AG~De~Bruyn, Ronald Nijboer, John Swinbank, Richard Fallows, et~al.
\newblock Lofar: The low-frequency array.
\newblock {\em Astronomy \& Astrophysics}, 556:A2, 2013.

\end{thebibliography}
\end{document}